\documentclass[
amsmath,amssymb,amsfonts,
aps,prb,
longbibliography,showpacs,
reprint,
floatfix
]{revtex4-1}

\usepackage{graphicx}
\usepackage{verbatim}
\usepackage{amsmath,amssymb,amsfonts}
\usepackage{cancel}
\usepackage{verbatim}
\usepackage{ifthen}

\usepackage{natbib}

\usepackage{graphicx} 
\usepackage{lastpage}

\usepackage{color}

\newcommand{\mum}{{\mu}\text{m}}

\begin{document}


\title{Flow-Driven Formation of Solid-like Microsphere Heaps}

\author{Carlos P. Ortiz, Robert Riehn and Karen E. Daniels}
\affiliation{Dept.\ of Physics, 2401 Stinson Dr., North Carolina State University, Raleigh, NC, 27695, USA.}

\date{\today}

\begin{abstract}
We observe the formation of heaps of repulsive microspheres, created by flowing a colloidal microsphere suspension towards a flat-topped ridge placed within a quasi two-dimensional microfluidic channel. This configuration allows for both shear and normal forces on the microspheres in contact with the ridge. The heaps, which form against the ridge, are characterized by two distinct phases: a solid-like bulk phase in the interior, and a highly-fluctuating, liquid-like state which exists along its leading edge. We observe that heaps only form above a critical flow velocity, \(v_c\), and that they are destroyed by thermal rearrangements when the flow ceases. We monitor the dynamics of heap formation using fluorescence video microscopy, measuring the heap volume and the angle of repose in response to microsphere deposition and erosion processes. We find that the steady-state angle of repose, \(\theta_f\), increases as a function of inflow velocity, \(v_\infty\), with a functional form \(\theta_f \propto \sqrt{v_\infty - v_c}\)\,. 
\end{abstract}


\maketitle

\section{Introduction}
Flowing particles will accumulate against an obstacle in their path, a phenomenon exhibited by industrial or agricultural grains accumulating in conical piles and microscopic particles undergoing membrane filtration, particularly ultrafiltration.\cite{Belfort1994} 
This similarity in behaviors has raised the question of whether there is an underlying commonality which explains the general phenomenon, despite the large difference in length scales, the vastly different importance of Brownian motion and kinetic energy, and the near-absence of interparticle friction in colloidal interactions.

The quest to understand the formation of piles of granular materials has, in fact, become a model system in which to investigate how bulk behaviors arise from highly cooperative effects at the grain scale. Although the basic mechanism for understanding the angle of repose has long been known,\cite{Coulomb1776} interest in sandpiles was rekindled by the pioneering work of \citet{Bak1987}. Since then, detailed studies of pile-formation have led to a quantitative understanding of the intermittency of granular dynamics near the threshold of flow,\cite{Frette1996, Abate2007} force chains,\cite{Geng2001,Liu1995} how local dynamic rearrangements\cite{Staron2002,Dauchot2010,Katsuragi2010,Nordstrom2011} contribute to pile instability, the prediction of avalanches,\cite{Ramos2009} memory effects,\cite{Geng2001}  frictional rheology in submerged flows,\cite{Boyer2011,Cassar2005} and the relationship between internal friction and the cessation of flow.\cite{Midi-2004-DGF} 

In this paper, we report on the formation of sandpile-like structures in a two-dimensional (2D), approximately hard-sphere, colloidal system by flowing a dilute suspension against a ridge within a microfluidic channel.  Unlike granular piles that are formed by gravity, the flowing colloidal suspension is driven by a static pressure gradient which partitions the system into coexisting gas-like, liquid-like, and solid-like microsphere phases.
We find that for sufficiently large P\'eclet number (corresponding to large flow velocities), stable heaps of microspheres are able to form by the deposition of particles out of a dilute suspension. When the flow is reduced, the heap re-evaporates in a thermally activated, reversible process.  Below, we quantify the conditions under which these heaps form, and report their key properties as a function of the flow velocity. 

Our work extends earlier work on dense 2D colloidal systems in which a complex phase diagram,\cite{Murray1989,Nugent2007} the influence of hard and soft interparticle potentials,\cite{Marcus1997,Zahn1999} the onset of layering,\cite{Pansu1983}, the dynamic response to supercooling,\cite{Grier1994} and the effect of confinement\cite{Bubeck1999} were demonstrated. Analogous granular experiments showed a phase space that included even a colloid-like ordered, fluctuating phase.\cite{Olafsen1998} Here we add the action of a combined compressive and shear field to the classical 2D colloid.  In  dense bulk suspensions the application of similar compressive gravitational fields\cite{Pusey1986a,Davis1989} and shearing field\cite{Ackerson1981,Robbins1991} have driven gas-fluid, fluid-solid and solid-solid transitions. 
Our system joins a growing range of techniques that utilize the tunability of key suspension parameters \cite{Yethiraj2007} (P\'eclet number, polydispersity, interaction potential) to explore the interplay between Brownian, steric, and specific interparticle potential effects.


\section{Experimental Setup}

\begin{figure}
\centering
\includegraphics[width=\columnwidth]{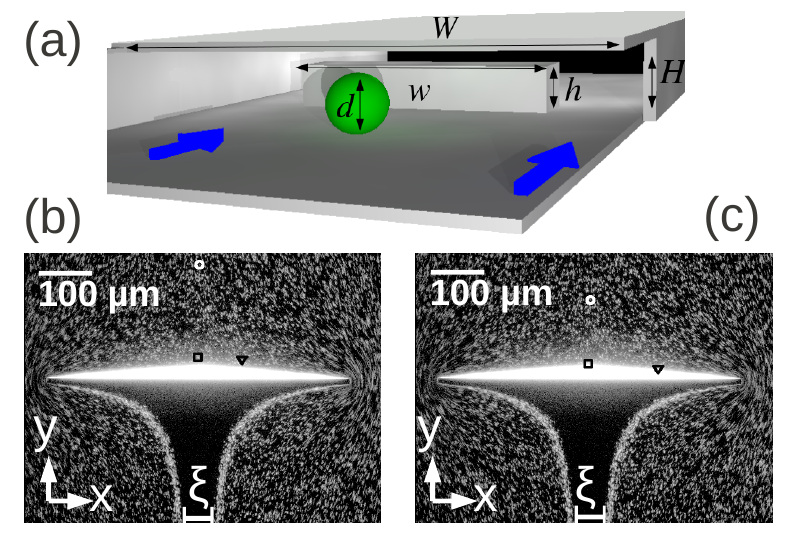}
\caption{(a) Schematic of microchip geometry (not to scale), showing a single microsphere flowing towards the flat-topped ridge over which fluid can flow. (b,c) Raw fluorescence images with linear grayscale (black = background, white = highest fluorescence intensity) of heap at steady state, at areal density \(\rho = 235/(100\mum)^2\) and inflow velocity $v_\infty = 33~\mum/s$. In the two seconds between (b) and (c), three tracked microspheres marked by the circle, square, and triangle symbols are shown in both figures to illustrate their relative displacements. }
\label{fig:expsetup}
\end{figure}

We perform experiments with a dilute suspension of fluorescent $d = 520$\,nm (5.9\% polydispersity) polystyrene microspheres (Bangs Laboratories) at concentration $0.1~\%~(w/w)$, corresponding to $6.7\times10^{11}$~microspheres/mL, suspended in a buffer solution described below. The suspension is pumped through a shallow microfluidic device with a flat-topped ridge rising from the channel floor. The full length of the microchannel is 14~mm, long enough to hydrodynamically decouple the access holes from the ridge.  The ridge runs perpendicular to the mean flow direction so that microspheres accumulate upstream of it (see Fig.~\ref{fig:expsetup}). The device channel has a quasi-2D geometry, with width $W = 200\,d$ and height $H = 1.8~d$.  The ridge has width $w = 100~d$ and occupies the central $50~\%$ of the channel width, permitting flow around the sides, and a height $h = 1.5~d$, permitting flow over the top. 
Because $d>H-h$, the streamlines which intersect the ridge push  microspheres against it with both normal and shear forces(see Fig.~\ref{fig:expsetup} b and c). Crucially, the geometry is resilient to catastrophic clogging since neither the ridge nor the resulting microsphere heaps span the entire width microchannel. 

The microfluidic channels are manufactured on a silicon wafer, using photolithography to define the channels and reactive ion etching to transfer the pattern into the wafer.  Access holes $\approx 300\,\mu$m in diameter are sandblasted through the silicon wafer with $17\,\mu$m alumina particles. The fluid system is closed by bonding the silicon wafer to a borosilicate coverglass (Corning) coated with a $20\,\mu$m layer of cross-linked PDMS. This hybrid PDMS-glass coverglass allows for reversible chip bonding, allowing many experiments to be performed on the same chip.\cite{Inglis2004,Inglis2010} The finished device is mounted to a sample holder that provides $10$-$\mu$L reservoirs to supply suspensions to the access holes. The suspension flow rate through the channel is set by a digital pressure regulator (AirCom PRE1-UA1) that applies pressurized air ($0-10$~kPa) into the o-ring sealed reservoir. 

To introduce the suspension into the chip without microspheres interacting with air-water interfaces, we first wet the chip with a microsphere-free buffer. Addition of 1\,$\mu$M sulforhodamine dye to the buffer allows verification of complete wetting of the devices by fluorescence microscopy, especially of the gap above the ridge. We then remove $\approx90\%$ of the wetting buffer from the reservoir and inject the microsphere-bearing suspension into the remaining buffer. This procedure ensures air bubbles are not introduced into the access hole, and thereby allows us to maintain a stable flow rate over the duration of the experiments. At the start of each experimental run, we flow the suspension into the device. 

Using fluorescence microscopy, we both record the local microsphere distribution at the ridge, and continuously determine the two free suspension parameters: inflow velocity $v_\infty$ and mean 2D areal density $\rho$, given by the number of microspheres per unit area. To do this, we use microspheres that are fluorescently labeled with DragonGreen dye ($480$~nm excitation, $520$~nm emission, Bangs Laboratories).  We obtain the inflow velocity $v_\infty$ by finding the angle of maximum autocorrelation in two-dimensional graphs of intensity as a function of time and vertical position for each image column. In the Hele-Shaw geometry, the lateral velocity field far from the heap is spatially uniform,\cite{NDarntonetal2001} except within a distance $H$ of the channel edges. Thus measurement of \(v_\infty\) over \(47~\mum \times 100~\mum\) centered \(200~\mum\) upstream from the ridge suffices to fully characterize the bulk flow. We use the same spatial region to measure $\rho$ by counting the number of extended maxima above a noise threshold. 

Experiments are performed on a Nikon Eclipse 80i fluorescence microscope, where the microsphere fluorescence is excited by a continuous-wave broadband  Xenon lamp (X-cite 120). Emitted light is collected with a 10X Plan Fluor objective, and $658 \times 496 \,\mathrm{\mum}^2$ images are acquired with an Andor Luca EMCCD camera. All images presented are oriented such that the mean inflow and outflow directions are vertical. The physical pixel size is $10\times10\,\mathrm{\mum}^2$, while the diffraction limited projection of a point object on the CCD at our emission wavelength is  $8\times8\,\mathrm{\mum}^2$.  This is larger than an equivalent microsphere diameter, and we are thus not able to resolve individual microspheres in dense piles. 
The total depth of field is $5.6\,\mu$m, larger than the depth of the microchannel. In the absence of external driving, we measure the diffusion constant of a dilute set of microspheres in the microchannel, far from the ridge, to be \(D=0.55\mathrm\,{\mum}^2/s\). Therefore, the microsphere diffusive time scale is $\tau_D \equiv d^2/D =0.49$\,s in the microchannel, so we set our camera's exposure time to $\tau_e=0.1$\,s. This exposure time is sufficiently short to observe microsphere thermal rearrangements because $\tau_e < \tau_D$.  As we will discuss below, we measured a lowered and non-constant diffusion coefficient due to wall-drag\cite{Sharma2010} which lead to increased actual $\tau_D$. 

For the microspheres to have reversible contacts, it is necessary to reduce attractive interactions. We use a combination of steric and electrostatic stabilization. For steric stabilization, we add a non-ionic surfactant \(0.1~\%~(v/v)\), equivalently 1.6~mM, Triton X-100 (Sigma Aldrich) to a stock suspension of concentration \(0.1~\%w\), which is enough surfactant to reach full coverage of the microspheres if the surface area per surfactant molecule is approximately \(12\,\AA^2\). To ensure that the surfactant coverage of the microsphere surface approaches full coverage, we tumble the suspension over 2 hours at \(\approx 10\)\,rpm to disrupt the surfactant micelles (CMC for Triton X-100 \(\approx 0.02~\%~(v/v)\), equivalently 0.22-0.24~mM). To test that we have successfully disrupted the micelles into covering the microspheres, we measure the particle size distribution of the tumbled solution using dynamic light scattering. While the presence of free surfactant introduces an entropic attractive force, the depletion force, we observe that its magnitude is not sufficient to form weakly aggregated gels over months in the absence of fluid flow, much less in the presence of flow. Thus, for our purposes, this depletion force is negligible.

We introduce a long-range repulsive electrostatic interparticle interaction by using microspheres with sulfate-modified surfaces ($\textrm{pKa} \approx 2.5$).  The interaction strength can be tuned via the buffer component of the suspension. We set $\mathrm{pH} = 5.4$ with a $10$~mM citric acid buffer in water titrated with sodium hydroxide. Using the Henderson-Hasselbalch equation, we estimate the concentration of free ions and calculate a Debye length of $~3$\,nm. Experimentally, the ionic strength of the solution is adjusted by dilution  of a high-concentration buffer with  deionized water until the conductivity  is $0.2\pm0.02$~mS/cm. The microsphere \(\zeta\)-potential is $\approx -43$~mV, ensuring the electrostatic stabilization of the suspension.\cite{W.B.RusselD.A.Saville1992}  All conductivity, \(\zeta\)-potential, and DLS measurements are measured with a Zetasizer (Malvern). 

\section{Results} 

\begin{figure}
\centering
\includegraphics[width=\columnwidth]{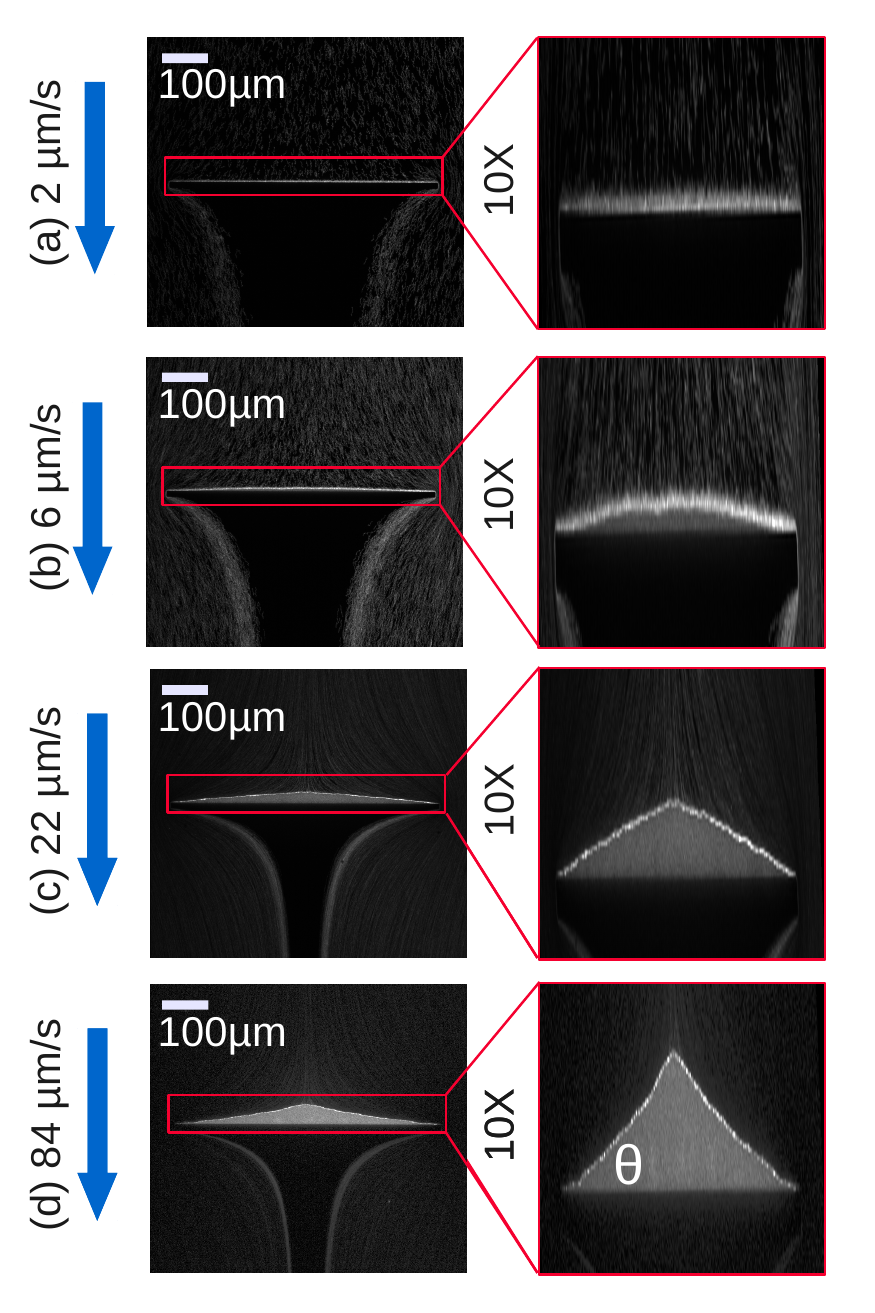}
\caption{Maps of mean density fluctuation amplitude of microsphere heaps formed at inflow velocities 
(a) $v_\infty = 2$~${\mu}$m/s, 
(b) $v_\infty = 6$~${\mu}$m/s, 
(c) $v_\infty = 22$~${\mu}$m/s, and 
(d) $v_\infty = 84$~${\mu}$m/s.  The bulk flow is from top to bottom.  Each pixel is given a grayscale value proportional to the standard deviation measured over $8$~s (80 images) after steady state was reached. Images in the left column have the equal horizontal and vertical scales, images in the  right column are stretched by a factor of 10 in the vertical direction. Bright areas at the surface of each heap are strongly fluctuating (fluid-like), while dark areas have a low fluctuation amplitude (solid-like). }
\label{fig:stdheaps}
\end{figure}

An illustrative image of a heap at steady state is shown in Fig.~\ref{fig:expsetup}bc. The inflow (top to bottom) carries microspheres towards the ridge that runs horizontally. Above the ridge, a dense, approximately triangular zone forms. We measure the geometry of the dense region by finding the points of steepest intensity gradient for each pixel column of the image, and taking the distance between them as the local extent $e(x,t)$ for each column,  where $x$ is the horizontal position and $t$ is the time. From each $e(x,t)$ profile, we compute the heap area by integrating $A(t) = \int e(x,t) \, dx $. An angle of repose $\theta(t)$ was defined from the slope of $e(x,t)$ at the outer edge of the ridge.

Although the suspension flow pattern around the ridge is disturbed by its presence, the microsphere-excluded zone below the ridge provides a clear indication of amount of liquid passing through the dense zone and over the barrier. We characterize the magnitude of this flow by the asymptotic physical extent of the excluded zone $\xi$ (see Fig.~\ref{fig:expsetup}b).  In Fig.~\ref{fig:expsetup}bc, the high microsphere areal density indicated by the bright bands delimiting the excluded zone is primarily the result of deterministic lateral displacement\cite{Huang2004} of the streamlines colliding with the heap. While the rate of microsphere erosion from the heap also contributes to an enhanced $\rho$, this rate is small at steady state.

We varied the incoming flow velocity over a range of $v_\infty = 1 - 100\,\mathit{\mu m/s}$ to explore the nature of the dense heap formed against the ridge. In Fig.~\ref{fig:stdheaps}, we plot the pixel-level fluctuation amplitude of the fluorescence intensity over 0.8~s (80 frames) after a steady-state heap was formed  (see below for definition of steady state). Under this analysis, microsphere-free regions and static microspheres appear dark. Moving, isolated microspheres appear as streaks which blend to a homogeneous gray at high flow velocities. The dense region appears gray, and the gray scale indicates the local magnitude of microsphere rearrangements (without correction for the local $\rho$). In all images, we observe a highly fluctuating bright layer  on the upstream side of the dense particle region.  For sufficiently high flow velocities, we observe a region of low fluctuation amplitude between the ridge and the bright layer (Fig.~\ref{fig:stdheaps}bcd). This region is not present at low flow velocities (Fig.~\ref{fig:stdheaps}a).  We interpret the presence of the low-fluctuating region as the formation of a quasi-2D heap, by analogy with granular heaps. In the following, we will refer to the region of low fluctuation amplitude as \emph{solid-like} and the region of high fluctuation amplitude as \emph{fluid-like}.  Note that the fluid-like phase appears to be compressed by the applied flow field in a similar way that an atmosphere can be compressed, but that the large $\rho$ makes it unlikely that the particles can be treated as a classical gas.

We observe that the heaps in Fig.~\ref{fig:stdheaps} do not form perfect triangles, even in the steady state. Instead, we find heaps with rounded peaks and convex sides at low $v_\infty$  (Fig.~\ref{fig:stdheaps}bc), and pointed tips and concave sides at high $v_\infty$ (Fig.~\ref{fig:stdheaps}d). We also observed a small degree of asymmetry between the left and right side of the heap.

\begin{figure}
\centering
\includegraphics[width=\columnwidth]{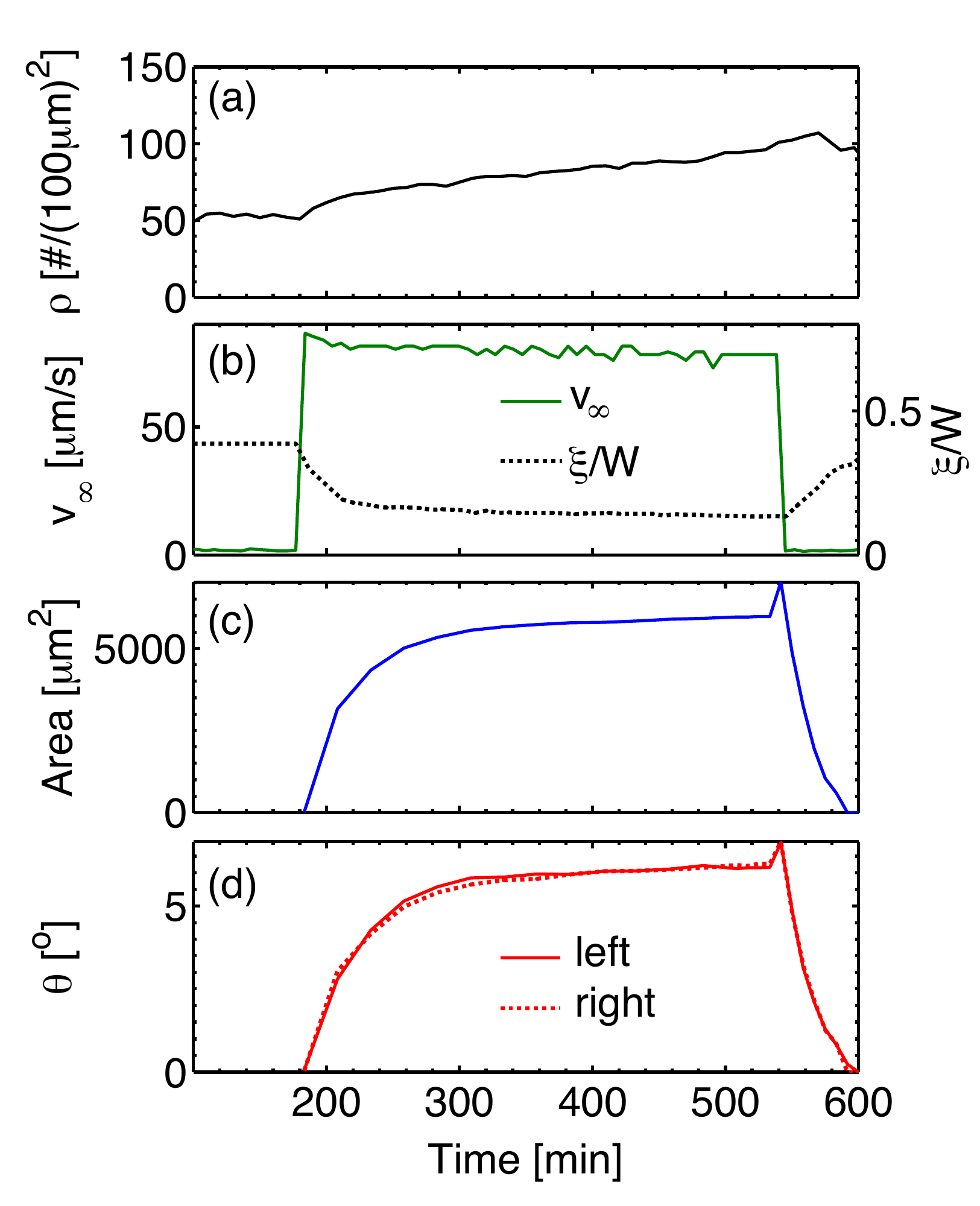}
\caption{Time-series of (a) microsphere concentration $\rho$, (b) inflow velocity $v_\infty$ and scaled width of excluded zone $\xi/W$ (c) heap area $A$, and (d) angle of repose $\theta$ in response to a six hour pulse of increased velocity. See Electronic Supplementary Information for a video of these dynamics. }
\label{fig:heapgrowth}
\end{figure}

To define the process and time scale over which the heaps reach steady-state, we perform experiments in which we form heaps with solid-like cores at constant $v_\infty$, and subsequently drop $v_\infty$ to a value at which only the fluid-like layer was previously observed. The results are shown in Fig.~\ref{fig:heapgrowth}. Initially, the device was primed with suspension at $v_\infty = 2\,\mu$m/s for 3 hours without heap formation. After $v_\infty$ was quickly increased to $84\,\mu$m/s, we observed an asymptotic increase of the heap area to $A\approx 6000$~$\mu$m$^2$  and a steady-state angle of repose of $\theta_f=6.2^\circ$.  The final heap contained $\approx 2 \times 10^4$ particles and was $\approx 56\,d$ tall at its peak. When $v_\infty$ was returned to $2\,\mu$m/s, a short-lived expansion of the heap was recorded.  Thereafter we observed a gradual reduction of $A$ and $\theta$ until only the fluid-like layer remained.  The process thus appears as if the solid-like phase gradually erodes, in a process akin to evaporation or melting. All stages of the experiment indicated that Brownian motion of microspheres is essential: an atmosphere is present on the forming heap at all times, the heap impulsively expands when $v_\infty$ is reduced, and the heap gradually reduces thereafter.  Visual inspection of the steady state dynamics and of the heap formation and erosion process are available through the Electronic Supplementary Information.

For this experimental run, we determined a deposition time constant of $42$~min and an erosion time constant of $17$~min by fitting exponential relaxation laws to $A(t)$ at early times (Fig.~\ref{fig:heapgrowth}c); similar values are obtained for $\theta(t)$. At longer times, we observe a non-exponential, gradual increase of $A(t)$ and $\theta(t)$; we believe this may result from the slow increase of the microsphere areal density throughout the high-$v_\infty$ phase from  $\rho = 50/(100\mum)^2$ to nearly twice that. Note that this corresponds to a change in the interparticle spacing of the suspension from $16 \, d$ to $11 \, d$, indicating that we remain in a weakly interacting gas-like flow regime during this time. We observed that variations within this range did not have a strong effect on the steady state, but did not conduct a detailed investigation of the effect of $\rho$ on the dynamics or final state.


\begin{figure}
\centering
\includegraphics[width=0.9\columnwidth]{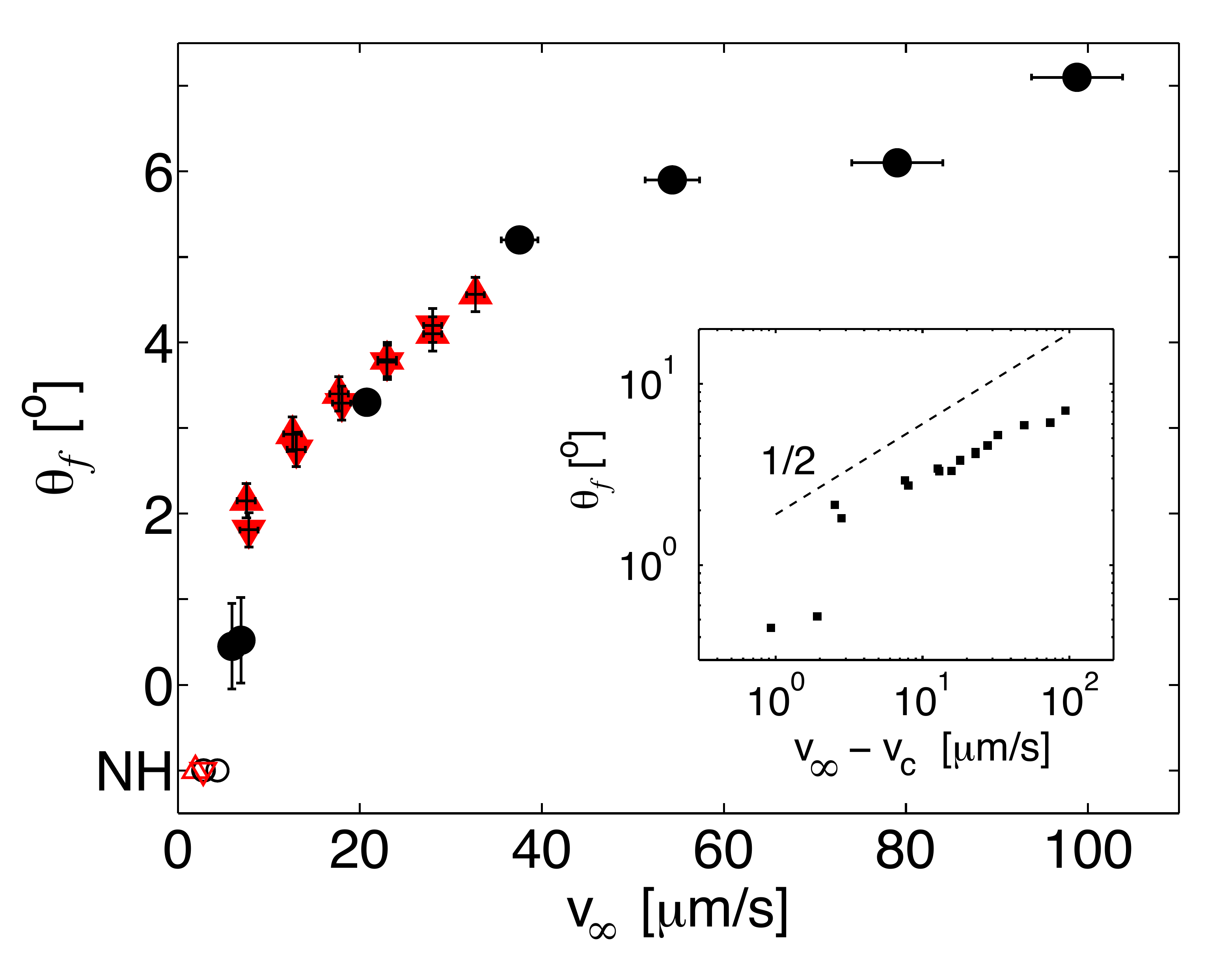}
\caption{Steady state angle of repose $\theta_f$ as a function of inflow velocity $v_\infty$. Solid points denote experiments in which a heap was formed and an angle could be measured; open symbols  denote experiments in which no heap (NH) was formed. Velocity error bars represent our measurement uncertainty in the mean measured, while the angle error bars represent the min/max range of angles observed at steady state. Red points correspond to the increase/decrease sweep shown in Fig.~\ref{fig:staircase}, with $\blacktriangle$ indicating the runs in which $v_\infty$ was increasing and $\blacktriangledown$ indicating decreasing. Inset: All solid (heap-forming) points plotted on logarithmic axes as a function of $v_\infty - v_c$, where $v_c = 5 \, \mu$m/s. Dashed line provides a comparison to a power law with exponent $1/2$.}
\label{fig:AnglesVvel}
\end{figure}
 
Fig.~\ref{fig:AnglesVvel} shows a detailed exploration of the functional relationship between heap geometry and $v_\infty$, as illustrated by the images shown in Fig.~\ref{fig:stdheaps}. At each value of $v_\infty$, we allowed the heap to reach a steady state, defined by a change in $A$ and $\theta$ smaller than $1\%$/hour (durations ranged from $2-6$ hours). To distinguish the cross-over to a heap-forming state, we observe that the two experiments with $\theta_f \approx 0.5^\circ$ are unambiguously heap-forming based on visual identification of the characteristic low-fluctuation amplitude region using the temporal standard deviation technique. In this regime, $\theta$ is difficult to measure accurately due to the small height of the heap ($\approx 2$ pixels). 

We observe that there is a well-defined relationship between $v_\infty$ and $\theta_f$, with finite-sized piles forming for values of $v_\infty$ above a critical velocity $v_c \approx 5\,\mu$m/s.  Note that Fig.~\ref{fig:AnglesVvel} is derived from two distinct experimental series with two different instances of device assembly, and that $\rho$ varied but was kept within the dilute regime. For the black points in Fig.~\ref{fig:AnglesVvel}, $\rho$ ranges from 40 to 150 \(/\mum^2\), while for the red points, $\rho$ ranges from 290 to 370\(/\mum^2\). The relative order of $v_\infty$ values was changed between the runs, and the possibility of hysteresis was checked via a series of pressure changes in which $v_\infty$ was systematically increased and then decreased to the same values. This case will be discussed in more detail below. We observe that the data, in spite of the variation in $\rho$, approximately collapse to a single curve which take the approximate form $\theta_f \propto \sqrt{v_\infty - v_c}$.

\begin{figure}
\centering
\includegraphics[width=0.8\columnwidth]{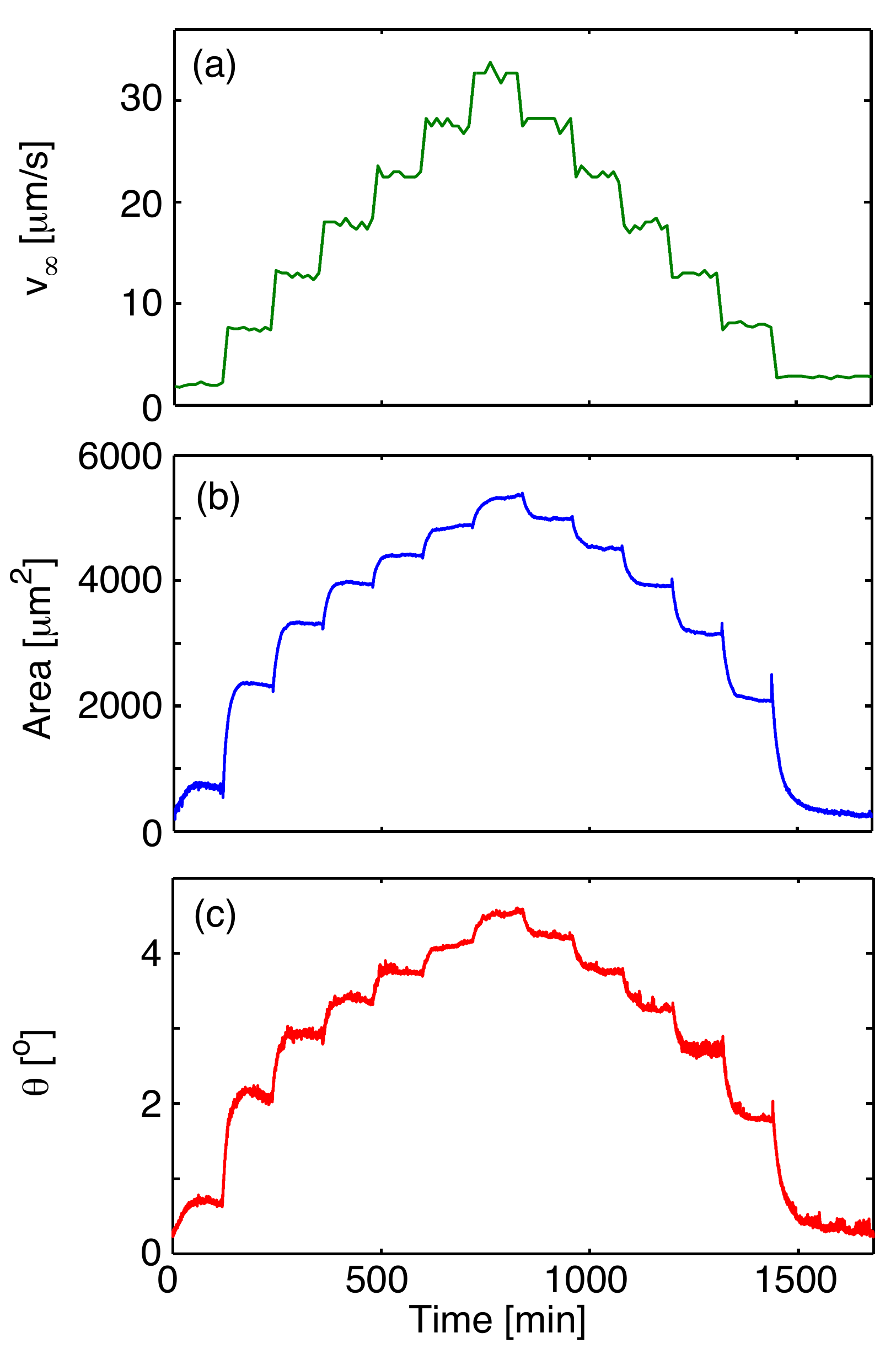}
\caption{Time-series of (a) inflow velocity $v_\infty$, (b) heap area $A$, (c) angle of repose $\theta$ for a series of increasing and decreasing pressure steps. The areal density for this run is $\rho = (330 \pm 42)/(100\mum)^2$, where the range covers two standard deviations. }
\label{fig:staircase}
\end{figure}

The detailed results of the hysteresis-probing experiment are shown in Fig.~\ref{fig:staircase}. The applied pressure is increased and decreased along a staircase profile, resulting in a velocity step of $6$~$\mu$m/s every two hours (see Fig.~\ref{fig:staircase}a), sufficient to reach steady state. The resulting $A(t)$ and $\theta(t)$ are shown in Fig.~\ref{fig:staircase}bc, and the triangular points in Fig.~\ref{fig:AnglesVvel} are derived from these plots. We observe only weak hysteresis, indicating that the steady state geometry is independent of whether driving is increasing or decreasing. We further note that the apparent hysteresis is correlated to the variation in $\rho$.

An interesting feature of  Fig.~\ref{fig:staircase} is the presence of sudden variations in $\theta$ (see Fig.~\ref{fig:staircase}b), which are not present in the pile area curve (Fig.~\ref{fig:staircase}c). Visual inspection of the raw video reveals that these likely represent rearrangement events on the heap surface that affect only the profile shape, but not the profile area. Interestingly, these rearrangements are more prominent during the second half of the experiment (erosion) than during the first half (deposition).

\section{Discussion}

There are significant similarities between granular heaps created by the raining of particles under gravity, and the microsphere heaps created in the present study: both are largely triangular in cross-section and can be built with different angles of repose. However, granular materials have an angle of repose which is determined by the static friction coefficient $\mu$ of the particle material,\cite{Coulomb1776} with larger angles supporting greater tangential forces through larger values of $\mu$. 
For example, \citet{Lemieux2000} observed $\theta \approx 20^\circ$ (equivalent to $\mu = \tan \theta = 0.36$) for heaps 
formed from dry, millimeter-scale, spherical, glass beads.


Compared to granular piles, we measure much lower angles of repose from $0.5^\circ$ to $7^\circ$ (see Fig.~\ref{fig:AnglesVvel}), and have no expectation of frictional contacts between the microspheres. Therefore, a closer comparison lies with simulations and experiments on frictionless particles. \citet{Peyneau2008} observed lower limit of $\theta \approx 5.8^\circ$ for a molecular dynamics simulations of a frictionless pack under shear, independent of the applied pressure. Similarly, for $120 \, \mu$m bubbles rising against to form an upside-down heap, \citet{Lespiat2011} observed a lower limit of $\theta \approx 4.6^\circ$. While these values are consistent with our observations, and support the expectation that friction play a minimal role, we find that we are able to form steady-state heaps with even lower angles of repose, approaching $\theta_f \approx 0^\circ$ (see Fig.~\ref{fig:AnglesVvel}). This suggests that at low P\'eclet number, Brownian microspheres are able to  explore their cages sufficiently well\cite{W.B.RusselD.A.Saville1992,Weeks2000} to destroy the geometrical constraints which give rise to the finite angle of repose in frictionless granular/bubble systems.  The fluctuations also lead to lubrication and frictionless contacts since even small thermal displacements are able to disrupt the contact network.\cite{Ancey2001}


We have described the microsphere heaps formed above $v_c$ as having a solid-like core, which we identified by its small temporal density fluctuations when compared to the dense region formed below $v_c$, or the boundary layer between the heap and the surrounding dilute suspension (Fig.~\ref{fig:stdheaps}). Whether the microspheres exist in a solid-like or liquid-like state depends on a competition between the stabilizing normal component of the drag force of the fluid flowing through the heap, its de-stabilizing tangential component, and the de-stabilizing Brownian motion of microspheres.  At the lowest inflow velocities (such as in Fig.~\ref{fig:stdheaps}a), the Brownian motion dominates and we observe only a liquid-like phase, without forming a heap. 
The liquid-like region of our system mirrors the density increase (concentration polarization) that is observed at the interface of a filter membrane in cross-flow filtration systems.\cite{Porter1972a} At sufficiently high suspension concentrations or flow velocity through the membrane, these systems also show the formation of a  filter cake, mirroring the solid-like bulk phase of the interior of the heap that we observe at high $v_\infty$ (Fig.~\ref{fig:stdheaps}bcd). Note that cross-flow filtration differs from our experiment in that the ratio of normal to tangential flow components is far larger in our system. It is commonly assumed that the cake phase is solid {\it per se}, while our heap still undergoes thermal reorganization.


\citet{Song1995} describe a similar fluid-like layer as having an exponential density distribution. They defined a criterion for cake formation based on the filtration number $N_F$, defined as the ratio of the energy required to bring a particle from the surface of the cake to the interface of bulk solution and concentration polarized layer. 
Our measurements support a similar criterion, in which we examine the relative importance of Brownian motion and fluid drag for the transition from a solely-liquid-like state (Fig.~\ref{fig:stdheaps}a) to one in which both phases are present (Fig.~\ref{fig:stdheaps}bcd).
We quantify the importance of Brownian motion with the P\'eclet number $\textrm{Pe} \equiv \tau_D/\tau_a$, where  $\tau_a = d/v$ is the particle advection time and $\tau_D=d^2/D$ is the self-diffusion constant $D$.  We measured 
$D=0.25 \,\mu\mathit{m}^2/\mathit{s}$ in a dense suspension that was created by ceasing the flow around a liquid-like state. 
A simple argument demanding flow continuity through the heap and over the ridge yields that the 
 mean normal velocity of the aqueous suspension at the shear boundary for particles is $v_\mathrm{heap} \simeq v_\infty\frac\xi w$.   Using our measured value of $D$, we obtain  $1 < \textrm{Pe} < 23$ in the incoming dilute suspension flow for the range of $v_\infty$ explored here. The critical velocity  $v_c \approx 5$~$\mu$m/s  corresponds to $\textrm{Pe} \approx 2$. This argument is only an estimate, as it neglects the influence of both shear and the incident particle density, both of which likely  influence $\theta_f$. In addition, the diffusion constant is modified by geometry- and concentration-dependent contributions to the Stokes drag \cite{Acrivos1987, Acrivos1992} as well as the degree of confinement. \cite{Weeks2000, Sharma2010}  Nonetheless, in experiments on dense suspension flows, \citet{Isa2009} also observed a change in rheological properties at similar values of  $\textrm{Pe}$.

The microfluidically-assembled microsphere heaps will also provide a convenient system in which to examine the behavior of a dense, solid-like colloidal system as a function of the applied stress.  In Figs.~\ref{fig:heapgrowth} and \ref{fig:staircase}  we note a transient compression and expansion of the heap upon sudden increase and decrease of the compressive stress set by $v_\infty$, respectively.  For instance, at $t = 533$~min in Fig.~\ref{fig:heapgrowth}cd the area heap transiently expands by 18\% under a drop in pressure of $\approx30\,Pa$, which corresponds to a modulus of $\approx10^2$\,Pa, far smaller than the few $10^9$ Pa typical for polystyrene. This observation provides evidence that individual microspheres are not in constant contact with each other but instead continually exploring cages by Brownian motion.

Finally we note that the experimental system presented here lends itself for repeated analysis of non-steady state systems that may be driven into supercritical regimes. For instance, we are able to create patches of colloids close to space-filling packing by releasing close-packed heaps through sudden stopping of the flow. As such, this system will be useful for exploring the temperature-dependence of jammed solids near the jamming transition.\cite{Liu-2010-JTM,VanHecke2010} Our observation of a critical $v_c$ above which jamming occurs is consistent with suggestions\cite{Trappe2001} that thermal systems should jam by increasing the osmotic or hydrostatic pressure. While not explored in the present study, the interparticle potential is tunable by adjusting the $\textrm{pH}$ of the buffer component of the suspension.

\section{Conclusion}

In this Paper, we have established a new system for forming a solid-like heap of microspheres within a microfluidic flow. We observe that the inflow velocity is the key control parameter in setting whether a steady-state heap is formed (large $v_\infty$, small P\'eclet number) or destroyed (small $v_\infty$, large P\'eclet number).  The steady-state angle of repose of this heap increases from $\theta_f \approx 0^\circ$ near the transition, to larger angles (up to $7^\circ$) at the fastest velocities explored, with a functional form $ \theta_f \propto \sqrt{v_\infty - v_c}$. We propose that this effect arises through the suppression of thermal fluctuations at larger flow velocities through reaching of a critical density of microsphere in a region of concentration polarization.


\section{Acknowledgements} 

The authors are grateful to James Gilchrist and Michael Shearer for stimulating conversations, and to John Chavez for technical assistance with the experiments.
Parts of this work were performed at Oak Ridge National Laboratory and the Cornell NanoScale Facility, a member of the National Nanotechnology Infrastructure Network, which is supported by the National Science Foundation (Grant ECS-0335765). C.P.O. was supported by an NSF Graduate Fellowship, with additional support from NSF grants DMR-0644743 and DMS-0968258.


\providecommand*{\mcitethebibliography}{\thebibliography}
\csname @ifundefined\endcsname{endmcitethebibliography}
{\let\endmcitethebibliography\endthebibliography}{}

\end{document}